\documentclass[nonatbib]{article}

\usepackage{arxiv}

\usepackage[utf8]{inputenc} 
\usepackage[T1]{fontenc}    
\usepackage{hyperref}       
\usepackage{url}            
\usepackage{booktabs}       
\usepackage{amsfonts}       
\usepackage{nicefrac}       
\usepackage{microtype}      
\usepackage{lipsum}
\usepackage[numbers]{natbib}
\usepackage{multirow}
\usepackage{graphicx}
\usepackage{float}
\floatstyle{plaintop}
\newfloat{program}{thp}{lop}
\floatname{program}{Code}

\title{Deep-diving of Atlantic salmon (\textit{Salmo salar})  during
  their marine feeding migrations}
\shorttitle{Deep-diving of Atlantic salmon}

\author{
  Sigurður Már~Einarsson\thanks{Corresponding author:
    \texttt{sigurdur.mar.einarsson@hafogvatn.is}}, \hspace{0mm}
  Sigurður~Guðjónsson,
  Ingi Rúnar~Jónsson and
  Jóhannes~Guðbrandsson\\
  Marine and Freshwater Research Institute\\
  Skúlagata 4, 101 Reykjavík, Iceland\\
  Tel.: +354-5752000 \\
  Fax: +354-5752001 \\}
\shortauthor{Einarsson \textit{et al.}}
\date{\today}

\begin{document}
\maketitle

\begin{abstract}
Data from seven data storage tags recovered from Atlantic salmon marked as smolts were analyzed for depth movements and patterns of deep diving during the marine migration. 
The salmon mostly stayed at the surface and showed diurnal activity especially from autumn until spring. 
During the first months at sea the salmon stayed at shallower depths ($<100$ m). 
The salmon took short deep dives ($>100$ m), that were rare or absent during the first summer at sea but increased in frequency and duration especially in late winter.  
The maximum depth of the dives varied from 419 to 1187 m. 
Most of dives were short, ($<5$ hours) but could last up to 33 hours. 
The duration of dives increased in late winter until spring and the overall depth and maximum depth per dive increased exponentially over time. 
The initiation of the dives was more common in evenings and at night, suggesting nocturnal diving. 
We hypothesized that deep diving is related to feeding of salmon as mesopelagic fish can be important food for salmon during winter. 
\end{abstract}

\keywords{Atlantic salmon \and DST tags \and deep diving}

\section{Introduction}
\label{intro}
Research in this century has begun to elucidate many details of the marine portion of Atlantic salmon (\textit{Salmo salar}) life history and ecology. 
Smolts generally leave their natal rivers in the spring \citep{Otero2014}, enter the sea as post-smolts, typically move rapidly through the coastal areas and fjords before entering the open ocean \citep{Thorstad2011,Lacroix2004,Gudjonsson2005} and tend to stay close to surface (1-2m) with irregular dives to 6.5m depth \citep{Davidsen2008}.  
Substantial knowledge gaps remain, however. 
Once the post-smolts leave the coast for open water, their movements are inadequately understood. 
Available evidence indicates that they are pelagic and mainly found in the surface layers of the sea \citep{Holm2006}, staying at temperatures between 8-12$^\circ$C \citep{Friedland2000} and salinities $> 35$ psu \citep{Holm2003}. 
Smolts tagged with data storage tags in Iceland remained close to the surface most of the time and showed diurnal movements, occupying deeper areas during the day. They stayed in areas with estimated surface temperatures from 6-15$^\circ$C, encountering warmer temperatures during the summer \citep{Gudjonsson2015}. 
The migration routes of salmon to their feeding areas also remain inadequately understood. 
Salmon occupy vast areas in the Atlantic Ocean and are known to aggregate for feeding in a number of areas where mixed stock fisheries historically took place including Greenland, the Faroe Islands and the Norwegian Sea \citep{Chaput2012}. 
The seas south and east of Iceland are also important feeding areas for salmon from UK, Ireland and southern Europe \citep{Olafsson2015}. 

In response to low and declining marine survival rates of salmon \citep{Chaput2012}, it is of paramount importance to understand the factors that influence their marine survival throughout their range. 
In the past, knowledge of marine life history and movements of salmon relied mainly on data gathered from salmon catches in commercial fisheries and research cruises. 
In recent years, however, advances in the use of satellites to gather oceanographic data have proven useful along with the use of fish telemetry including data storage tags (DST), acoustic tracking and pop-up satellite archival tags (PSATs). 
Due to their small size DSTs have been especially effective in identifying and monitoring thermal preferences of salmon in coastal habitat \citep{Reddin2004,Reddin2006}.
DSTs and PSATs have been used to record thermal and depth preferences of salmon kelts \citep{Reddin2011,Chittenden2013,Lacroix2013,Strom2017}. 
DST technology is also particularly useful for detecting deep-diving, brief, rapid, regular or irregular movements that may be associated with thermoregulation or foraging, and other ecological benefits to salmon \citep{Reddin2006}. 
Similar benefits may apply to  other fish species; e.g., \textit{Mobula tarapacana} \citep{Thorrold2014}. 
From an  individual fish' short-term changes in depth and temperature, inferences can be made about general aspects of deep-diving in relation to temperature and other ecological factors.

In this study, we applied DST technology to monitor thermal and depth preferences of salmon tagged as smolts, released from, and recaptured in a river in southwest Iceland. 
In a previous analysis of data from these tag recoveries, \citet{Gudjonsson2015} used temperature profiles and diurnal activity to geolocate the DST tagged salmon.  
The identified diurnal activity pattern of the salmon was not surprising as it has been previously reported for Atlantic salmon and other salmonid species \citep{Reddin2011,Rikardsen2007,Walker2000} and may be a response due to vertical migrations of many organisms that salmon feed on like amhipods, squids and euphausids that move into the upper part of the water column at night \citep{Pearchy1984,Davis1998}.  
A more surprising behavior of the salmon was rare deep dives ($>100$ m) that have previously not been reported in detail for older salmon and never before for post-smolts.

The objective of this study was to analyze the frequency and duration of deep dives ($>100$ m) during the marine stay, and investigate the maximum depth displayed by the salmon. We investigated diel patterns in deep-diving and how the frequency, duration and depth of deep dives changed during the ocean migration as the smolts grew to adult salmon.

\section{Materials and Methods}
\label{MnM}
\subsection{Tagging and recovery of fish}
Smolts used in this study were hatchery-reared fish produced at the Laxeyri Fish Farm from multi-sea winter parents of the River Rangá Stock. 
In 2005 and 2006 we tagged 598 smolts (mean length, 19.0 cm $\pm$ 0.89 cm SD; mean weight 78.9 g $\pm$11.3 g SD) with DST-micro tags (Star-Oddi, Garðabær, Iceland). 
The tags (2.5 g in weight, 25.4 mm in length) were set to record temperature and pressure (depth) at hourly intervals. 
The tags were surgically inserted in the body cavity of anaesthetized smolts; a detailed description of the exact tagging procedures is provided in \citet{Gudjonsson2015}. 
The tagging of smolts were done in accordance with the relevant Icelandic legislation on animal welfare in collaboration with the Icelandic Food and Veterinary Authority.
Tagged smolts were released into the River Kiðafellsá (lat 64.314701$^\circ$N, lon 21.776101$^\circ$W) in southwestern Iceland, where they could migrate volitionally to the estuary and sea. 
From these releases, seven one-sea-winter salmon with DST tags were recovered (Table \ref{tab:WandL}), five in 2006, and two in 2007. 
All the salmon were recaptured in a fish trap were all returning salmon  were scanned for tags in River Kiðafellsá, except one fish (tag 331) caught by an angler in the neighboring river Elliðaár. 
More details about release and recaptures are given in \citet{Gudjonsson2015}.
For analysis of depth use and deep diving, complete datasets of hourly temperature and depth profiles were acquired for all seven fish, except for the fish caught in River Elliðaár where the tag stopped recording data on January 13th 2006.

\subsection{Analysis of movements among depths and inferences of deep-diving}
The recovered data were first inspected qualitatively to categorize depth measurements as deep or shallow.  
The shallow measurements mostly consisted of diurnal dives were the salmon were deeper during the day than the night but still at rather shallow depths (within 100m). 
The deep dives were to depths below 100m that in most cases were rather short. In very few cases we could not categorize the dive as shallow or deep. 
We suspect that since the resolution of our measurements was sparse we might have missed some deep dives, and the confusing dives represent a transition between the two patterns but missed the deepest point of the dive. These dives were categorized as intermediate.

We used a mixed linear model to test for changes in: depth, maximum depth and duration per dive over time. 
The individual was the random factor in the models. Both the slope and the intercept were allowed to vary between individuals. We also assumed first order auto-correlation of errors within individuals (AR1 process). The variables were log-transformed before model fitting. We fitted the models using maximum likelihood (ML) and tested for fixed effects, random effects and error structure with a likelihood ratio test. For parameter estimation we refitted the models with restricted maximum likelihood (REML). The calculations were performed using nlme-package \citep{Pinheiro2009} in R (see Code \ref{code1}).

$\chi^2$-goodness of fit was used to test for difference between time of day in dive onset frequency and total duration of dive event. 

\section{Results}
\label{Res}
\begin{figure*}[ht]
  \centering
  \includegraphics[width=\linewidth]{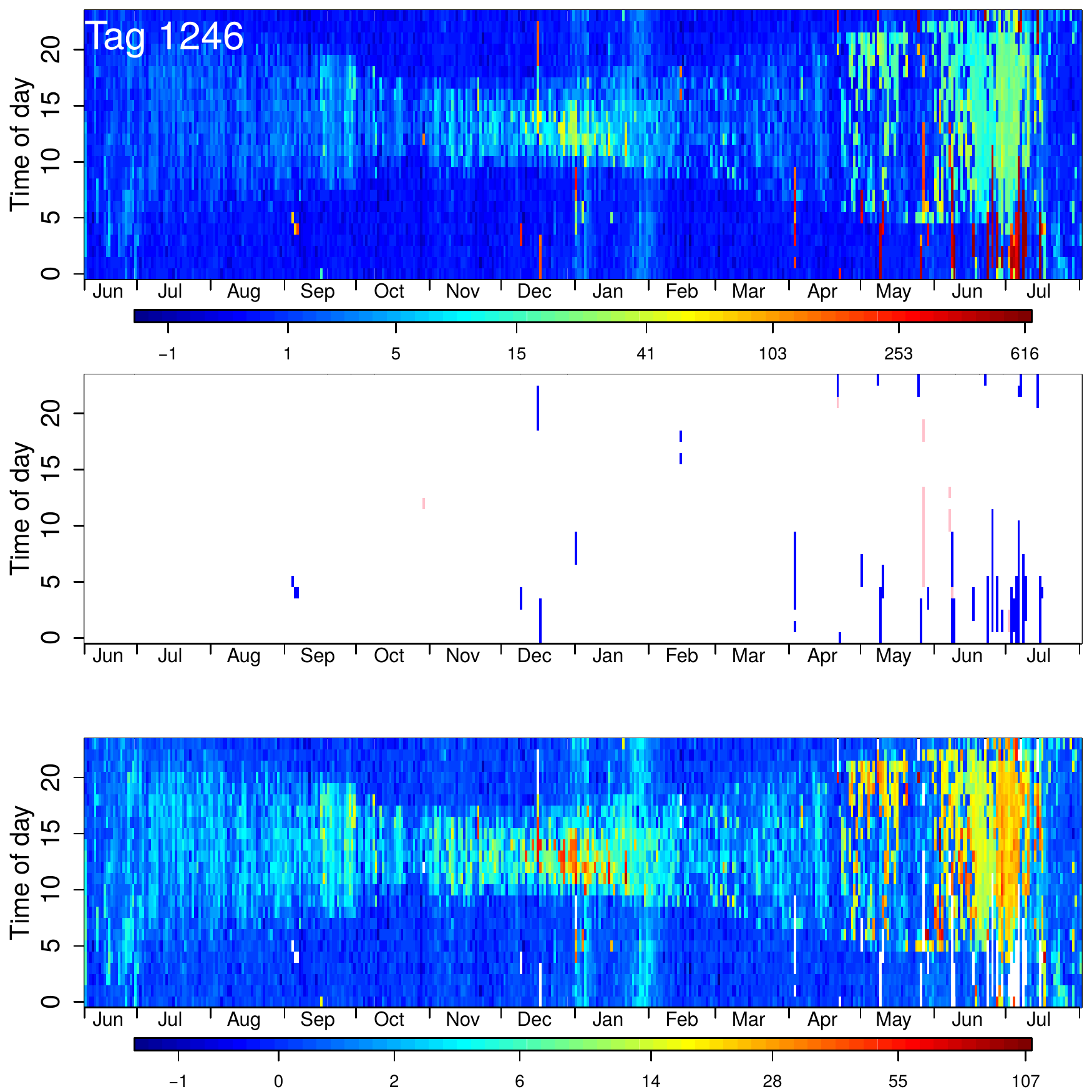}
  \caption{Depth profile at sea for tag 1246. First the complete profile is shown. In the middle panel the deep dives and intermediate measurements are highlighted (deep dives blue, intermediate behavior pink). 
The bottom panel gives the depth profile without deep dives, making the
shallow diel depth behavior more apparent. 
Color scale shows the depth in the top and bottom panels. Notice that the depth color scale is log-scaled. See supplementary figure \ref{fig:alldep} for comparable figure for all the tags.}
  \label{fig:heatmap}
\end{figure*}

Upon qualitative, visual inspection of the depth profiles recorded by the DST tags, two distinct patterns were evident. 
Firstly, all seven of the salmon showed distinct diurnal movements where salmon stayed at greater depth during the day and were nearer to the surface at night (Fig. \ref{fig:heatmap} and \ref{fig:alldep}). 
This diurnal pattern was clearly related to the day length (Fig. \ref{fig:heatmap} and \ref{fig:alldep}) and was much more pronounced in autumn, winter and spring, when days were shorter but in the long daylight hours of summer the diel pattern was not as evident. Analysis of the diel patterns indicated that in the first months of summer and autumn (post-smolt stage), the salmon occupied relatively shallow areas during the day, somewhat greater depths in winter and spring, but in all seasons remained almost entirely shallower than 100m.

\begin{figure*}[ht]
  \centering
  \includegraphics[width=\linewidth]{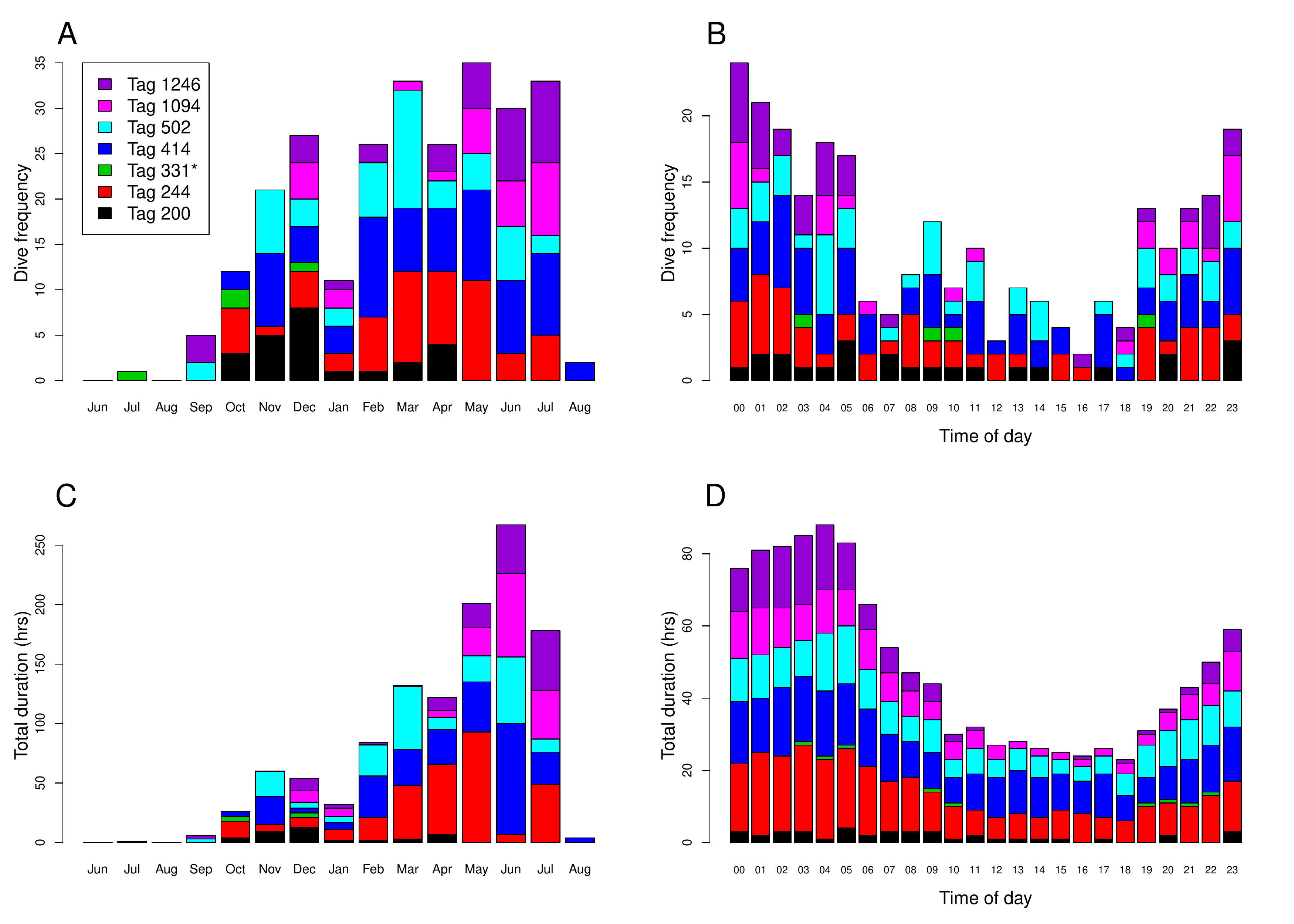}
  \caption{Deep dive frequency and total duration of deep dives by month (A and C) and time of day (B and D). The colors indicate different tags as the legend in A describes. * Tag 331 stopped measuring January 13.}
  \label{fig:freq}
\end{figure*}

\begin{figure*}[ht]
  \centering
  \includegraphics[width=\linewidth]{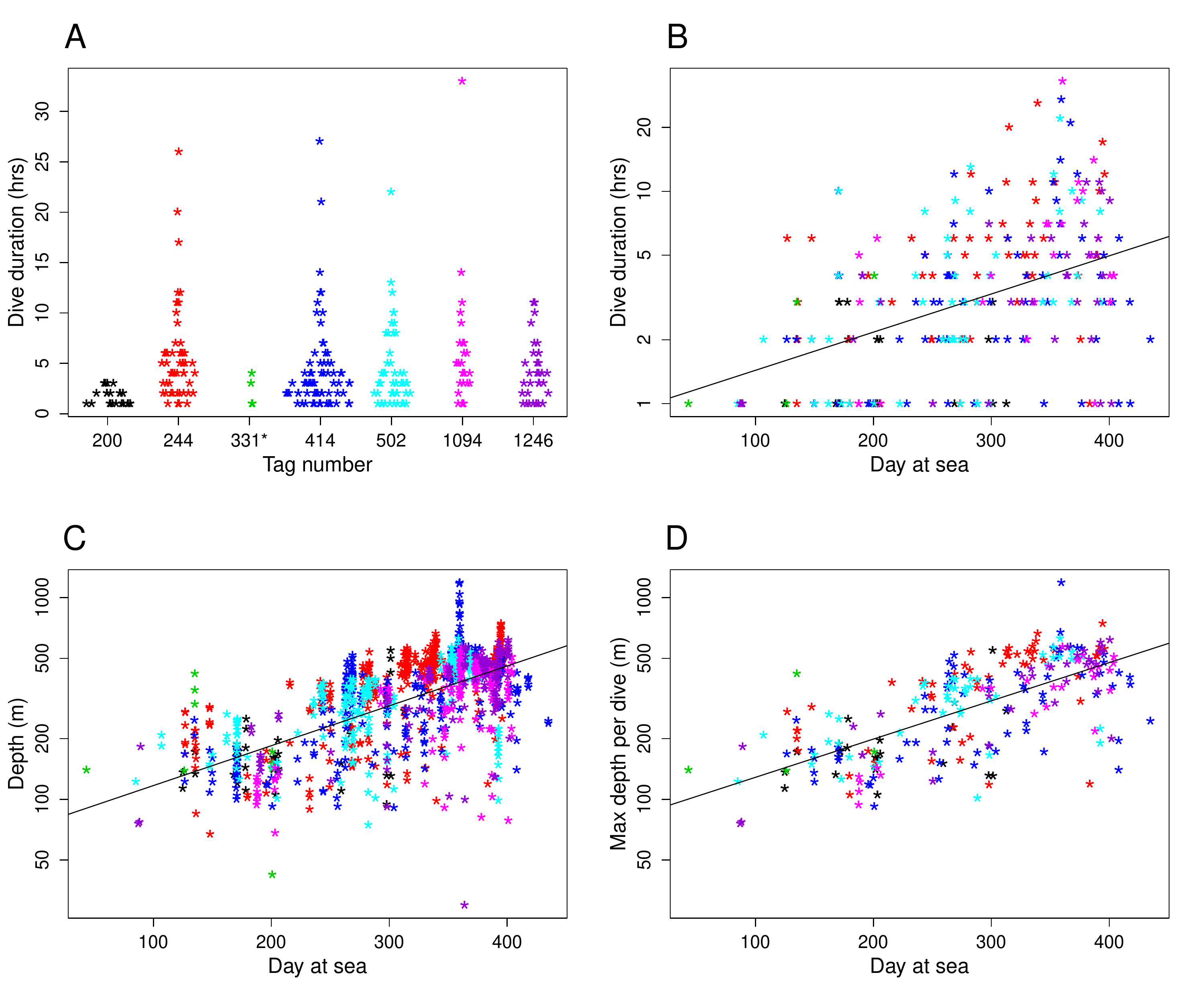}
  \caption{Dive duration and depth in deep dives by individual fish.
Panel A: Plot of all dive durations undertaken by individual fish (plotted as tag number); 
Panel B-D: Pooled data on individual dives from all fish are plotted with associated regression lines, to show trends over time in duration of dives (B), depth in dives (C) and maximum depth per dive (D).
Points are colored by fish (see A). 
See Table \ref{tab:mod} for the formula of regression lines. 
Notice that the y-axis is not linear in figures B, C and D. 
* Tag 331 stopped measuring January 13.}
  \label{fig:lendep}
\end{figure*}

Secondly we noticed the presence of occasional deep dives (below 100 m) that seemed to be independent of the shallow diurnal movement pattern described above. These dives were rare or absent during the first summer at sea but increased both in frequency and duration over time (Fig. \ref{fig:freq} A and C). 
The deep dives were in most cases of rather short duration (5 hours or less, see Fig. \ref{fig:lendep} A) and the median value for individual fish ranged from 1.5 - 5 hours (Table \ref{tab:lib}). 
There were individual differences in the display of this pattern both over time and overall (Fig \ref{fig:freq}A, C and Table \ref{tab:lib}). 
The duration of deep dives was between 0.41\% to 3.06\% of the total time at sea and between 0.04\% and 0.19\% of the time was spent in intermediary dives for each fish. (Table \ref{tab:lib}).

All the fish displayed deep dives and maximum depth of the dives ranged from 419 - 1187m (Table \ref{tab:lib}). 
Frequency of the onset of dives was significantly different by time of day (Fig. \ref{fig:freq}B, $p < 0.0001$, $\chi^2$-test); dives were more common in the evening and at night than during the day. 
The total duration of the dives also significantly differed by the time of day (Fig. \ref{fig:freq}D, $p < 0.001$, $\chi^2$-test). 
Duration of dives increased over time (Fig. \ref{fig:lendep}B, Table \ref{tab:mod}) as an individual fish's duration of marine residence increased and the fish grew. 
Most of the dives were of short duration (5 hours or less) and the minimum duration for all the fish was one hour (one measurement). 
Long dives, lasting more than 20 hours were displayed by four of the fish, the longest extending 33 hours (Table \ref{tab:lib}, Fig. \ref{fig:lendep}A and \ref{fig:alldep}). 
Both the overall depth and maximum depth per dive increased exponentially with increased time and growth of fish during their marine stay (Fig. \ref{fig:lendep}C and D, Table \ref{tab:mod}).  We did not find the individual variation in dive duration and depth to be significant. The measurements were on the other hand found to be significantly auto-correlated (Table \ref{tab:mod}).

\section{Discussion}
\label{Dis}
The diel movements and deep dives observed in all seven Atlantic salmon in this study is consistent with the scarce information available from other studies. 
\citet{Westerberg1982a} noted during tracking studies that salmon made dives to deeper areas. 
Diurnal depth movements in Newfoundland Atlantic salmon post-smolts were recorded by DST temperature tags during early marine life \citep{Reddin2006}, and additional DST recordings of depth and temperature we obtained for four adult salmon tagged in the Norwegian Sea in 2004 \citep{Holm2006}. 
For the latter fish, which were recaptured in home waters, the proportion of time spent at the surface decreased during daylight hours \citep{Holm2006}. 
Similar patterns were recorded by DST tags in Atlantic salmon kelts released and recovered in a Newfoundland river; these fish spent 18\% less time near the sea surface during the day than during the night \citep{Reddin2011}. 
Kelts tagged with Pop-up satellite archival tags (PSATs) in Miramichi River in Canada also spent most of the time in the surface layers and exhibited deep diving \citep{Strom2017}. 
PSATs are used externally on the fish and affect diving behavior as diving frequency, depth and diving speed are lower compared to fish tagged with internal tags \citep{Hedger2016}. 
\citet{Hedger2017} found that post-spawners monitored with PSATs from 3 salmon populations in the Norwegian and Barent Seas spent 82\% of their time near the surface ($<10$ m).
In other salmonids, DST technology have shown that Arctic char at sea demonstrated a diel diving pattern, staying deeper during the day \citep{Rikardsen2007} and temperature data from DST tags on four species of Pacific salmonids imply diel patterns of diurnal descents to deeper, cooler water followed by nighttime ascents to surface \citep{Walker2000}.

The smolts used in the study were of hatchery origin \citep{Gudjonsson2015}. 
Releases of Atlantic salmon hatchery smolts are well established in Iceland, both for ocean ranching and enhancement releases in rivers with poor natural production resulting in annual catches of thousands of fish in the angling sport fishery \citep{Thordardottir2017}.   
Hatchery smolts have generally lower returns than wild smolts \citep{ICES2018}, but usually show the same fluctuation in returns as wild salmon \citep{Johannsson1996} and generally seem to exhibit similar timing of river entry and behavior in upstream migration as wild salmon returning to the rivers.

Of particular interest in our study was the presence of rather rare short and deep dives (below 100 m) undertaken by all the salmon. 
Such deep dives were almost absent during the first months at sea, increased in frequency and duration thereafter, occurred mostly at the pre-adults and adult stages over the several months prior to their spawning migration especially in the period from January to June.  
Estimates of geolocation of the fish suggests the salmon mainly stayed in the Irminger Sea during this period \citep{Gudjonsson2015} in the vicinity of the Reykjanes Ridge where bottom depths range from $500 - 1,500$ m.  
Exact location of the fish within this area is not known, but it is unlikely that the bottom depth limits the diving depth of the salmon. 
\citet{Holm2006} observed deep dives down to depths of 280 m from four adult salmon tagged in the northeast Atlantic Ocean. 
Tagging of kelts have also revealed such dives \citep{Reddin2011,Chittenden2013,Strom2017}.

The causes of diel movements and deep diving observed in this study may be a result of several driving forces including thermo-regulation, predator avoidance and foraging.
Considering foraging, it is known that Atlantic salmon are opportunistic feeders, using a wide variety of available prey throughout their marine stay \citep{Rikardsen2011}. 
Many of their prey include organisms like euphausiids, amhipods and squids that all undertake migrations to the upper water column at night \citep{Walker2000} supporting the idea that the diel patterns are driven by trophic factors. 

The exact causes leading to deep dives are not clear. \citet{Westerberg1982a,Westerberg1982b} suggested that dives could be associated with homing as salmon dove to orient themselves according to currents. Many diverse arrays of cues for homing have been proposed including smell, magnetic compasses and infrasound patterns in the ocean. The exact mechanism of homing, and its relation to deep dives, if any, remain unknown \citep{Thorstad2011}.

Another potential cause of deep dives is predator avoidance. 
Predation on salmon in the open ocean is not well documented, but salmon have been found in a few cases in stomachs of sharks, skates, cod (\textit{Gadus morhua}) and halibut (\textit{Hippoglossus vulgaris}) \citep{Ward2011}. \citet{Lacroix2014} showed that kelts tagged with Pop-up tags were preyed on by porbeagle shark (\textit{Lamna nasus}) and bluefin tuna (\textit{Thunnus thynnus}). However, the salmon in our study did not undertake deep dives during their first months at sea when the fish was still small, and the need for predator avoidance is probably greater than at the pre-adult and adult stage when the fish is larger with more ability to avoid predators. Predator avoidance may not be a primary cause of deep diving in this case.

Deep diving and diel movements may also be related to foraging benefits.
An increase in diving depth ($> 200$m) was observed in adult Atlantic salmon in the Barents Sea monitored by DST and PSATs, in winter and spring, possibly indicating foraging \citep{Hedger2017}. 
Pre-adult and adult pre-spawning salmon in the northwest Atlantic Ocean feed mostly on different species of fish with capelin (\textit{Mallotus villosus}), sand eels (Ammodytidae), herring (Clupeidae), lantern fishes (Myctophidae) and barracudinas (Paralepidae) being the most common preys \citep{Rikardsen2011}. 
Pre-adult and adult salmon in the northeastern Atlantic Ocean feed more extensively on crustaceans although fish like lantern fish, barracudinas and herring are important prey during certain times of the year \citep{Rikardsen2011}. 
In the sea around the Faroe Islands, various crustaceans were important food items for salmon in the autumn but various mesopelagic fish as lantern fishes, pearlsides (Sternoptychidae) and barracudinas were eaten in the winter \citep{Jacobsen2001}. 
Analysis of stomach samples of Atlantic salmon caught in Icelandic waters from 2010-2015 as a bycatch in the mackerel (\textit{Scomber scombrus}) fishery has revealed that lancet fish (Paralepidae) was the most important food item in this period (unpublished data of the authors). 
Mesopelagic fishes such as lantern fishes and barracudinas are therefore known as important prey items for Atlantic salmon and could explain the nocturnal deep dives of salmon as they sought mesopelagic fishes which in turn were  following zooplankton migrations in the water column.
Recent results indicate that the biomass of mesopelagic fishes in oceanic ecosystems needs to be revised as their biomass might be up to 15,000 million tons \citep{Irigoien2014}. 
Diel and other vertical movements of prey and deep dives of predators may thus be related linking different ocean zones through the food web. 
For example, extreme cases of vertical movements and deep dives have been reported in rays \citep{Thorrold2014}. 
The life histories, movements, and behaviors of many mesopelagic fish species as well as their population sizes is poorly known. 

Additional research is needed to clarify the relationships of salmon to their marine habitats and food webs. In addition, in future years, more recoveries of DST-tagged fish will enable a stronger ecological and statistical evaluation of the depth usage and deep diving observed in this study. 

\section{Acknowlegements}
We thank D.L. Scarnecchia for a critical review of the manuscript and appreciate valuable comments from two anonymous reviwers. Sincere thanks to several coworkers at the Marine and Freshwater Research Institute, especially Björn Theódórsson, Halla Kjartansdóttir and Eydís Njarðardóttir for their valuable contribution.  
The study was partially supported by the Salmonid Enhancement Fund in
Iceland.

\bibliographystyle{unsrtnat}
\bibliography{DST}

\clearpage
\section*{Tables}

\begin{table}[ht]
  \centering
  \caption{Individual returns of DST tagged salmon in R. Kiðafellsá 2006 to 2007. 
    Gender labeled as F for female and M for male, 
    the tagging date (Tag. date) and length (L) and weight (W) of the time of tagging is shown.
    Release date (Rel. date), Recapture date (Rec. date) and length (L) and weight (W) at the time of recapture are tabulated. Reprinted with changes from \citet{Gudjonsson2015}.}
  \begin{tabular}{rcrrrrrrr}
    \hline
     \multirow{2}{*}{Tag Nr} & \multirow{2}{*}{Gender} &   \multicolumn{3}{c}{Tagging} & \multirow{2}{*}{Rel. date} &\multicolumn{3}{c}{Recaptures}\\
     \cline{3-5} \cline{7-9}
     && Tag. date & L (cm) & W (g) && Rec. date & L (cm)&W (g)  \\
    \hline
     200 & F & 31.5.2005 & 19.9 & 83.5&  7.6.2005 & 17.8.2006 &59  & 1811 \\
     244 & F & 31.5.2005 & 18.5 & 70.7&  7.6.2005 & 17.8.2006 &64  & 2316 \\
     331 & F & 31.5.2005 & 18.5 & 60.8&  7.6.2005 & 13.8.2006 &60 & 2070 \\
     414 & F & 20.5.2005 & 17.7 & 59.1&  7.6.2005 &  1.9.2006 &61.5& 2210 \\
     502 & M & 20.5.2005 & 18.7 & 70.2&  7.6.2005 & 17.8.2006 &70.5& 3162 \\
    1094 & M &  9.3.2006 & 20.2 &102.4& 31.5.2006 & 13.8.2007 &65  & 2756 \\
    1246 & F &  9.3.2006 & 18.7 & 68.7& 31.5.2006 & 3.8.2007 &59.3& 1890 \\
    \hline
    \multicolumn{3}{c}{Averages}&18.9&73.6&&&62.8&2316\\
    \hline
  \end{tabular}
  \label{tab:WandL}
\end{table}

\begin{table}[ht]
  \centering
  \caption{Basic statistics of deep dives for each tag. The number of days at sea, 
  the number of dives (Dives), 
  the total duration of dives (Tot. dur.),
  the total duration of intermediate behavior (Int. beh.),
  the median length of dives (Med. dur.),
  the longest dive (Max. dur.) 
  and the maximum depth (Max. depth) for each tag.
  * Tag 331 stopped measuring January 13 and is omitted in calculation of mean values.}
  \begin{tabular}{rrrrrrrr}
    \hline
    Tag Nr& Days at sea & Dives & Tot. dur. (hrs) & Int. beh. (hrs) & Med. dur. (hrs) & Max. dur. (hrs) & Max. depth (m)\\
    \hline
    200  & 406 & 24 &  40 & 11 &1.5&  3 & 546\\
    244  & 421 & 55 & 309 &  4 &5& 26 & 742\\
    331* & 219 &  4 &   9 &  0 &2&  4 & 419\\
    414  & 446 & 71 & 298 &  6 &3& 27 &1187\\
    502  & 408 & 48 & 212 &  6 &3& 22 & 627\\
    1094 & 424 & 26 & 159 &  8 &4.5& 33 & 571\\
    1246 & 419 & 34 & 140 & 19 &3.5& 11 & 616\\
    \hline
    Mean value & 421 & 43 & 193 & 9 & 3.5 & 20.3 & 714\\
    \hline
  \end{tabular}
  \label{tab:lib}
\end{table}

\begin{table}[ht]
  \centering
  \caption{Significance of explanatory variables and estimates of parameters for deep dive duration (Dur.), depth in dives (Dep.) and maximum depth per dive (Max.). The model was $log(y_{i,t}) = a_i + b_it +\varepsilon_{i,t}$ or $y_{i,t}=C_ie^{b_it}\cdot e^{\varepsilon_{i,t}}$.
Even though each individual $i$ had a specific slope and intercept we only show the overall parameters (fixed effects). The errors were also modeled to have 
an AR1 structure with parameter $\theta$. The R-code for this is available (see Code \ref{code1}).}
  \begin{tabular}{rlllrrr}
    \hline
    & Time & Individual & AR1 & $C$ & $b$ & $\hat{\theta}$ \\
    \hline
    Dur. & **** & $ns$ & *    & $0.94$ & $0.0042$ & $0.16$\\
    Dep. & **** & **   & **** & $77.7$ & $0.0044$ & $0.48$\\
    Max. & **** & $ns$ & **** & $86.5$ & $0.0042$ & $0.30$\\
    \hline
    \multicolumn{7}{l}{%
    \begin{minipage}{6.5cm}%
      \resizebox{.6\textwidth}{!}{\begin{tabular}{ll}
                                    $ns$ & not significant, $p > 0.05$\\
                                    $*$ & $p < 0.05$ \\
                                    $**$ & $p < 0.01$ \\
                                    $***$ & $p < 0.001$ \\
                                    $****$ & $p < 0.0001$
                                  \end{tabular}}
                              \end{minipage}%
                                             }\\
  \end{tabular}
  \label{tab:mod}
\end{table}

\clearpage
\section*{Figures}

\newpage
\clearpage
\section*{Supplementary material}

\setcounter{figure}{0}

\makeatletter 
\renewcommand{\thefigure}{S\@arabic\c@figure}
\renewcommand{\theprogram}{S\@arabic\c@program}
\makeatother

\begin{program}
\caption{R-code for fitting linear mixed model, running likelihood ratio tests and extracting parameter estimations.}
\begin{verbatim}
#das = Day At Sea
#Full model
m1 <- lme(log(d.org)~das, data=dep.tab,
          random=~das|ind, method="ML",
          correlation=corAR1(form=~1|ind))
#No autocorrelation
m0.0 <- lme(log(d.org)~das, data=dep.tab,
            random=~das|ind, method="ML")
#No random effect
m0.1 <- gls(log(d.org)~das, data=dep.tab,
            method="ML", 
            correlation=corAR1(form=~1|ind))
#No increase in time
m0.2 <- lme(log(d.org)~1, data=dep.tab,
            random=~1|ind, method="ML",
            correlation=corAR1(form=~1|ind))

anova(m1, m0.2)#Test for time effect
anova(m1, m0.0)#autocorrelation
anova(m1, m0.1)#individuals

update(m1, method="REML") #Refitting with REML
#Parameters y=a*exp(bt)
c(exp(fixef(m1)[1]),fixef(m1)[2])
cbind(exp(coef(m1)[,1]),coef(m1)[,2])
\end{verbatim}
\label{code1}
\end{program}

\begin{figure*}[ht]
  \centering
  \includegraphics[width=.85\linewidth]{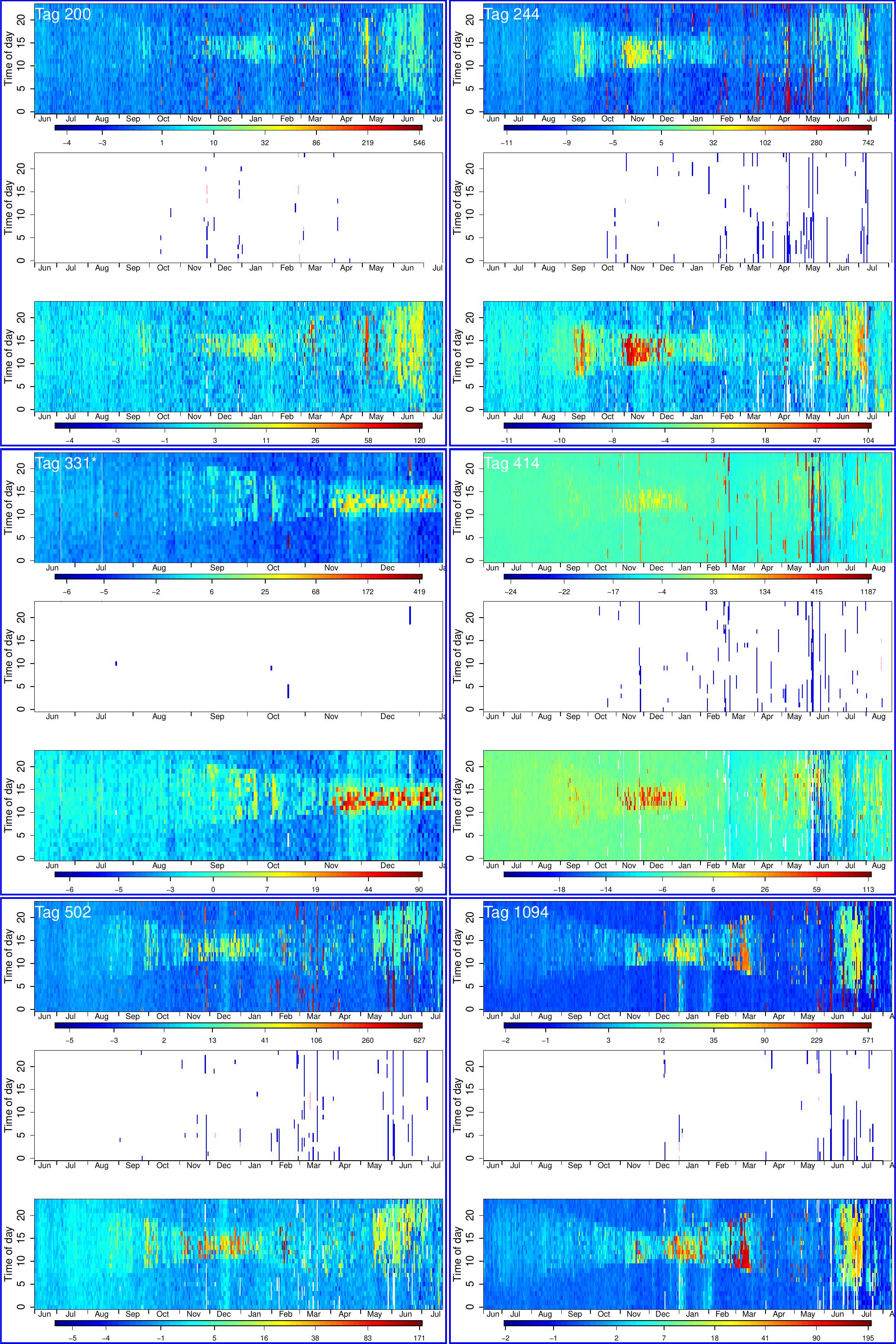}    
  \caption{Depth profile at sea for all tags except 1246 (see Figure \ref{fig:heatmap}). First the complete profile is shown. In the middle panel the deep dives and intermediate measurements are highlighted (deep dives blue, intermediate behavior pink). 
    The bottom panel gives the depth profile without deep dives, making the shallow diel depth behavior more apparent. 
    Color scale shows the depth in the top and bottom panels. Notice that the depth color scale is log-scaled.}
\label{fig:alldep}
\end{figure*}

\end{document}